# Multiple-Symbol Interleaved RS Codes and Two-Pass Decoding Algorithm

Z. Wang, A. Chini, M. Kilani, and J. Zhou

Broadcom Corp., California, 92617, USA

*Abstract* — For communication systems with heavy burst noise, an optimal Forward Error Correction (FEC) scheme is expected to have a large burst error correction capacity while simultaneously owning moderate random error correction capability. This letter presents a new FEC scheme based on multiple-symbol interleaved Reed-Solomon codes and an associated two-pass decoding algorithm. It is shown that the proposed multi-symbol interleaved coding scheme can achieve nearly twice as much as the burst error correction capability of conventional symbol-interleaved Reed-Solomon codes with the same code length and code rate.

## I. INTRODUCTION

TWO important error types present in most digital communication systems are random and burst errors. Random errors are typically the result of independent error events and are scattered. Burst errors are the result of correlated noise events and appear in clusters. In some applications, such as automotive networks with high level interference from adjacent systems, burst noise is much more prominent than random noise. In such situations, an optimal FEC scheme is expected to have a large Burst Error Correction Capability (BECC) while owning moderate Random Error Correction Capability (RECC) at the same time. Most of the existing FEC schemes are designed for heavy random errors that are inefficient in dealing with burst errors [1], [2].

Interleaved Reed-Solomon (IRS) codes were proposed when both large RECC and BECC are emphasized [3], [4]. The IRS codes integrate a high RECC of RS codes with interleaving techniques to improve BECC. An IRS code usually consists of several independent RS codes with the same code length. Depending on the codeword dimension, homogeneous and heterogeneous IRS codes were developed with their associated decoding algorithms, such as shift-register synthesis-based joint decoding method [5] and interpolation-based probabilistic decoding algorithm [6].

Existing IRS schemes interleave the independent RS coders at single-symbol level and adopt a single-pass decoding method. In this letter, we present a novel IRS code. The codewords are interleaved at the multiple-symbol level and a two-pass decoding algorithm is used. Without knowing burst

Authors are with Infrastructure & Networking Group, Broadcom Corp., California, 92617, USA. (e-mail: zfwang88@gmail.com).

error locations, the proposed Multiple Symbol IRS (MS-IRS) scheme (or burst-interleaved RS coding), can achieve nearly twice as much as the BECC of conventional Single Symbol IRS (SS-IRS) codes with the same code length and code rate. In addition, the BECC and processing latency of the proposed MS-IRS codes can be optimized by adjusting the length of each independent RS codes. A two-pass decoding algorithm is presented to enable the increased BECC of MS-IRS schemes. Specifically, in the first pass, one or more independent RS codes are decodable if the length of burst errors corrupting a specific RS code is smaller than the error correction capability of the code. An approximate burst error location can be derived from the corrected codewords in the first pass and then a second pass decoding will be performed with combined error and erasure coding [7], which improves the BECC.

This letter is organized as follows. Section II introduces the MS-IRS encoding scheme. Theoretical BECC and decoding latency are studied. Section III presents the proposed two-pass decoding algorithm. Monte-Carlo simulations on an example channel are presented in Section IV.

## II. MS-IRS CODES

Assume IRS codes with an interleaving depth equals to $L$. Each independent RS code is denoted by RS($N$, $K$, $m$), where $N$ is the number of symbols in a codeword, $K$ is the number of source data symbols in a codeword, and $m$ is the number of bits in each symbol. The burst-length (denoted as $BL$) refers to the number of symbols in each dispatch of data to one component code. Fig.1 illustrates the block diagram of a general IRS codes encoding scheme. A DEMUX splits the source data and feeds one of $L$ independent RS encoders at each instant. A MUX combines the encoders output back into one data stream.

Fig. 2 (a) shows an example of a conventional SS-IRS code with three independent RS codes and an interleaving depth of $L$ = 3. Assume each RS code has an error correction capability of four symbols. In addition, assume one symbol correction capability for random errors that are outside of burst noise. Now, we investigate the maximum burst error length when all independent RS codes are still decodable without knowing the burst error location. In this case, the maximum length of burst errors that can be corrected is three symbols. Fig. 2 (a) shows the maximum burst error length at the best and the worst cases, respectively, when all RS codes can be decoded correctly. For example, the burst noise at the best case can start from the first bit of the first code-1 symbol and end at the last bit of the third code-3 symbol. The burst noise at the worst case can start from the last bit of the first code-1 symbol and end at the last bit of the third code-3 symbol.

Fig. 2 (b) shows an example of an MS-IRS code also with $L$ = 3 and the length of each burst interleaving is 3-symbol (i.e., $BL$ = 3). Similar to Fig. 2 (a), the maximum burst noise length at the best and the worst cases are shown with one symbol error correction capability reserved for random errors. Adopting the proposed two-pass decoding algorithm, we need to decode at least one RS code correctly in the first pass. In the case shown in Fig. 2(b), the code-3 is decodable if the length of burst error is not too long. The burst noise at the best case can start from the first bit of the first code-1 symbol group and end at the last bit of the second code-2 symbol group. The burst noise in the worst case can start from the last bit of the first code-1 symbol group (or code-1 segment) and end at the last bit of the second code-2 symbol group.

In general, assume each RS code has an error correction capability of $t$ symbols over $GF(2^m)$. Without knowing the burst noise locations, the SS-IRS and MS-IRS codes have a BECC of $(L*t-1)*m+1$ bits and $(L-1)*2t*m+1$ bits at the worst case, respectively. It means that MS-IRS codes achieve nearly twice of the BECC of conventional SS-IRS codes when $L$ is large.

For MS-IRS codes, the above BECC equation assumes the number of symbols in each colored segment is equal to $t$, i.e., $BL=t$. If the number of symbols in each colored segment is less than $t$, a combination of random and burst errors can be corrected. Then, the BECC is equal to $(L-1)*2BL*m+1$ bits, in this case. In general, we choose $BL$ such that $2BL>>t>BL$ in order to achieve good tradeoff between BECC and RECC.

We now calculate the encoding and decoding latency of the MS-IRS codes. At the transmitter, a data buffer is needed to accommodate for increased data rate after FEC. The buffering latency is calculated based on the total FEC block and parity size. At the receiver, latency includes receiving time for the interleaved code plus decoding latency. For example, given a RS code (108, 96, $t$=6) over $GF(2^9)$ and 4X interleaving with $BL$ = 5, and assuming 1Gbps of data rate, the buffering latency is 4*(108-96)*96/108*9*1ns = 384ns. The receiving latency is 4*96*9*1ns = 3456ns. Decoding latency can be less than 120ns. Therefore, the total latency associated with FEC is less than 4μs.

## III. Two-Pass Decoding Algorithm

To achieve the increased BECC in the MS-IRS codes, we propose a two-pass signal decoding algorithm. Fig. 3 illustrates the flow diagram of the proposed two-pass signal decoding algorithm. In the first pass, perform RS decoding as usual and checks if at least one RS code can be decoded. Based on the decoding result, we can determine the burst error location and predict the erasure starting segment when at least one RS code is decodable. In this case, erasure decoding (or combined error and erasure decoding [7] when burst length $BL$ is smaller than error correction length $t$) will be performed in the second pass of error decoding. Therefore, a longer burst of errors can be corrected.

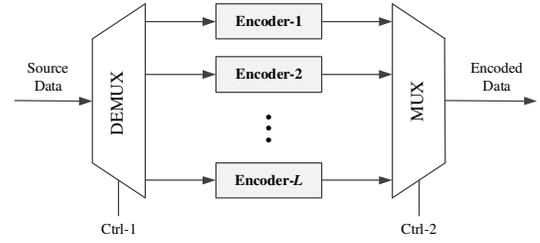

Fig. 1. Block diagram of a general IRS codes encoding scheme.

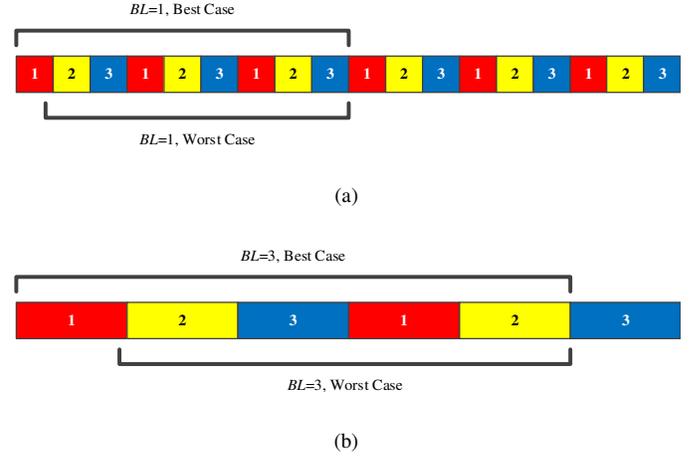

Fig. 2. An illustration of the maximum burst error length at the best and worst cases when all codes can be decoded correctly. Interleaved depth is 3. (a) SS-IRS codes, $BL$=1; (b) MS-IRS codes $BL$=3 with two-pass decoding algorithm.

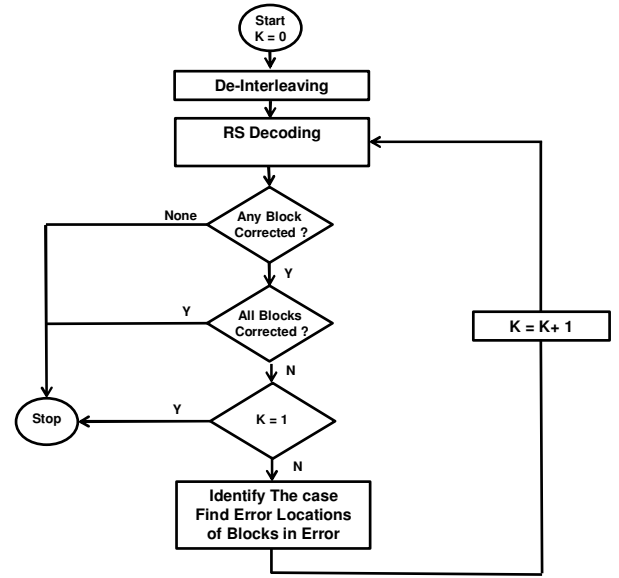

Fig. 3. Flow chart of the two-pass MS-IRS codes decoding algorithm.

## IV. Simulations

In this section, the advantage of the proposed MS-IRS code is verified by simulating an example communication system with PAM3 modulation scheme. The block diagram of the system model is shown in Fig. 4. The short RS code blocks are burst interleaved in the interleaver block, whereas for the long RS code, the interleaver is not applied. The output of the RS encoder and the interleaver is applied to a mapper that maps the symbols (3 bits each) to physical layer values {-1, 0, +1}. The method of mapping is given in Table I. The physical layer symbols {-1, 0, +1} are then transmitted.

The channel is assumed to be real valued and the imaginary and the real parts of symbols at the mapper output are transmitted as the even and odd samples. The channel response is depicted in Fig. 5. It is seen that the channel is dispersed over several physical layer symbols, which results in severe Inter Symbol Interference (ISI). To remove the ISI, a Decision Feedback Equalizer (DFE) is used at the receiver. The block diagram of the channel, noises, and the receiver equalizer (DFE) are shown in Fig.6. Note that it is assumed that the channel response is known to the receiver. The slicing is performed in one dimensional form as soon as one symbol is received; it is compared against a threshold (±0.5) and accordingly is selected from the set {-1,0,-1}. Once a pair of even and odd are collected, the symbol bits are obtained by demapper (see Table I).

Normally the RS decoder is able to correct the errors during the burst duration. In practice, however, the number of errors is more than the number of symbols corrupted by the burst noise. This is due to the error propagation problem associated with the DFE in which when the slicer makes a few decisions, the errors propagate through the feedback filter and arrive at the slicer input causing more errors to occur. These errors are bursty in nature and could easily extend to a number beyond the error correction capability of RS codes. In the following simulation cases, it will be shown that short RS codes, when followed by a multi-symbol interleaver, significantly improve the performance when compared with long RS codes.

### A. Case 1

This case compares a long code RS($N$=432, $K$=387, $t$=22, $m$=9) $L$=1, with a short code RS($N$=144, $K$=129, $t$=7, $m$=9) that has been multi-symbol interleaved with parameters $L$=3 and $BL$=6. The channel noises are assumed to be AWGN with Signal-to-Noise Ratio (SNR) of 30dB and a burst noise. The burst duration and period are set equal to 38 symbols and 5400 symbols, respectively. The Bit Error Rate (BER) and Block Error Rate vs. the Signal to Burst Ratio (SBR) are shown in Fig.7 and Fig.8, respectively. It is seen that the short RS code with multi-symbol interleaver with single pass exhibits a very similar performance to the long RS code. But, it is clear that RS(144,129) with a multi-symbol interleaver ($L$=3, $BL$=6) when used with a two-pass decoder performs significantly better than RS(432,387)

### B. Case 2

In this case, RS($N$=147, $K$=132, $t$=7, $m$=9), $L$=3, $BL$=7, is compared against RS($N$=144, $K$=129, $t$=7, $m$=9), $L$=3, $BL$=6, under the same channel conditions as given in Case 1, except that the burst duration is increased to 114 symbols. The BER and Block Error Rate vs. the SBR are shown in Fig.9 and Fig. 10, respectively. It is seen that RS($N$=147, $K$=132, $t$=7, $m$=9), $L$=3, $BL$=7 performs better than RS($N$=144, $K$=129, $t$=7, $m$=9), $L$=3, $BL$=6. In other words, depending on the burst duration and severity of the burst error, the $BL$ value can be selected in such a way that the best performance can be achieved without necessarily increasing the RS code length and latency.

| Symbol Bits | 000 | 001 | 010 | 011 |
|---|---|---|---|---|
| Mapper Output {even, odd} | {-1,-1} | {-1,0} | {-1,+1} | {0,-1} |
| Symbol Bits | 100 | 101 | 110 | 111 |
| Mapper Output {even, odd} | {0,+1} | {+1,-1} | {+1,0} | {+1,+1} |

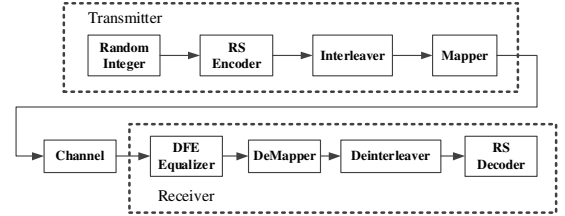

Fig. 4. Simulation model block diagram.

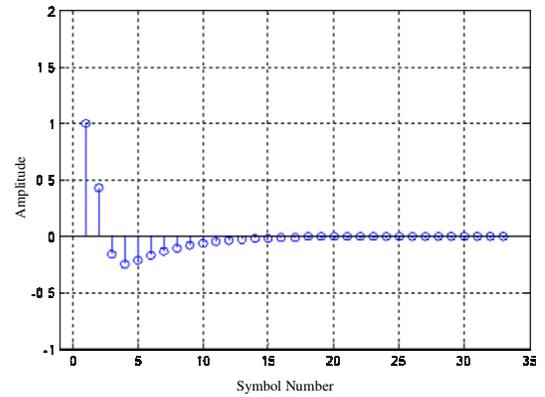

Fig.5. Channel response.

## V. CONCLUSIONS

In this letter, we have introduced the multi-symbol interleaving scheme, specifically multi-symbol interleaved RS coding, together with a two-pass decoding algorithm. The guidance about selecting interleaving parameters is given, and the detailed simulation results demonstrated the benefits of the proposed coding method. It should be noted that the component codes can also be BCH, LDPC, or other FEC codes.

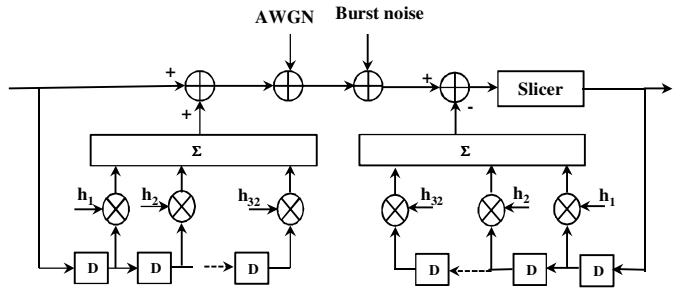

Fig. 6. The channel and receiver Equalizer block diagram.

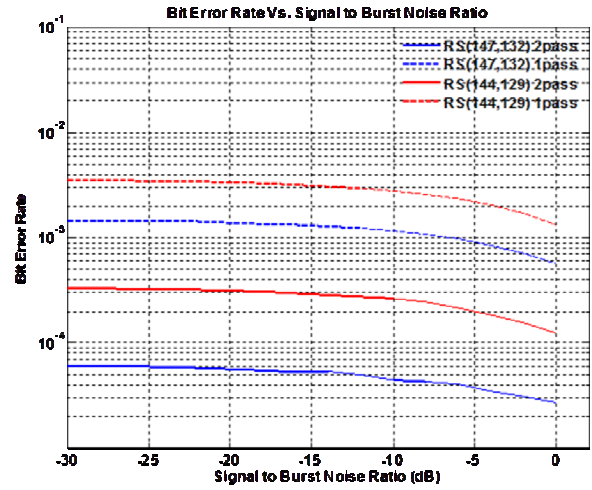

Fig. 9. Bit Error Rate vs. Signal to Burst Noise Ratio for short RS code but different burst length (Blue line *BL*=7, Red line *BL*=6).

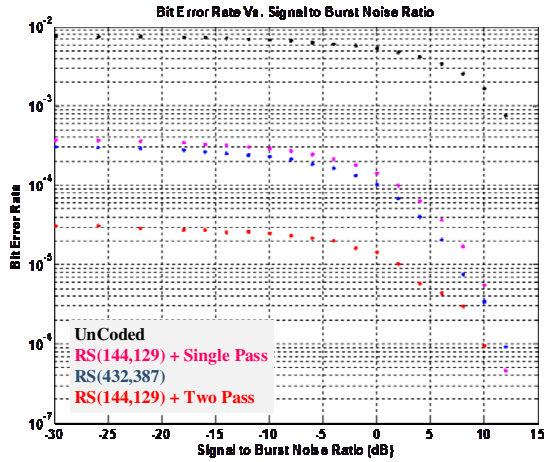

Fig. 7. Bit Error Rate vs. Signal to Burst Noise Ratio (Case 1).

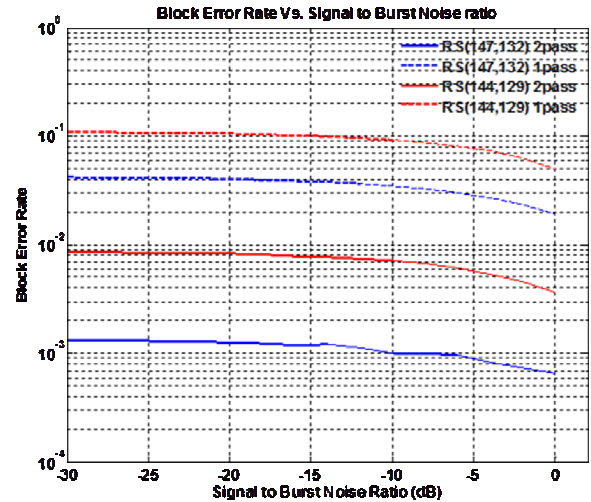

Fig. 10. Block Error Rate vs. Signal to Burst Noise Ratio for short RS code but different burst length (Blue line *BL*=7, Red line *BL*=6).

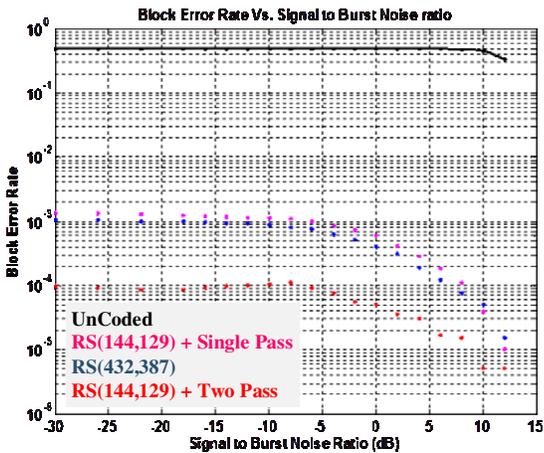

Fig. 8. Block Error Rate vs. Signal to Burst Noise Ratio (Case 1).